# The Origin Magnetars[1]

## ── The role of anisotropic neutron superfluid of neutron stars

Qiu-he Peng ( qhpeng@nju.edu.cn )
（Department of Astronomy, Nanjing University, Nanjing, 210093, China）

Abstract： We estimate the strength of the induced magnetic field due to the total induced magnetic moment of the anisotropic ($^3PF_2$) neutron superfluid interior of neutron stars. The induced magnetic field of the neutron stars is found as follows. $B^{(in)} = \frac{\eta}{T_7 - \eta} B^{(0)}, \quad \eta = \frac{m(^3P_2)}{0.1 m_{Sun}} R_{NS,6}^{-3}$

($T_7$ denotes the interior temperature of the neutron star in unit of $10^7$ K). The induced magnetic field will gradually increase with the temperature of the neutron star decreasing in their late evolutionary stage. A magnetar may appear in a condition when $T_7 \to \eta$. The upper limit of the induced magnetic field of $^3P_2$ superfluid is $3.35 \times 10^{14}$ η.

## 1. A Question

It is generally believed that there is a very strong magnetic field, B > $10^{12}$ gauss, for most of the neutron stars (e.g. Shapiro and Teukolski, 1984). There probably are magnetars with ultra-strong magnetic field strength over the quantum critical threshold, $H_{cr}$ =4.414×$10^{13}$ gauss (Duncan & Thompson 1992; Paczynski;1992; Usov 1992; Thompson & Duncan, 1995,1996). Anomalous X-ray Pulsars (AXPs) and Soft Gamma Repeaters (SGRs) are classes of candidates to magnitars (e.g. Kouvelliotou et al. 1998, 1999; Hurley et al. 1999; Mereghetti&stella 1995; Wilson et al. 1999).The magnetic field of the magnetars may be such strong as $10^{14}$-$10^{15}$ gauss.

What is the origin of the ultra-strong magnetic field for the magnetars? It is a very interesting question. Some models for origin of magnetars have been proposed. Ferrario & Wickrammasinghe (2005) suggest that the extra-strong magnetic field of the magnetars is descended from their stellar progenitor with high magnetic field core. Iwazaki(2005) proposed the huge magnetic field of the magnetars is some color ferromagnetism of quark matter. Recently Vink & Kuiper (2006) suggest that the magnetars originate from rapid rotating proto-neutron stars.

We are to propose a new model of origin of magnetars in terms of the behavior of the anisotropic $^3PF_2$ neutron superfluid in this paper.

## II  On anisotropic $^3P_2$ neutron superfluid

There are two kinds of neutron superfluid interior of a neutron star in general: one is the isotropic $^1S_0$ neutron superfluid with critical temperature of phase transition $T_\lambda \approx 1.0\times10^{10}$K.

---

[1] This research is supported by Chinese National Science Foundation No.10573011, No.10273006, and the Doctoral Program Foundation of State Education Commission of China

The density range of the isotropic $^1S_0$ superfluid neutron state is $\rho \sim (1.0\times10^{11}\text{-}1.6\times10^{14})$ g/cm$^3$..

Another superfluid is the anisotropic $^3P_2$ neutron superfluid. Energy gap calculations show that the pairing gap varies with the density of neutrons ( see Fig. 8 of Elgagøy et al. (1996)) as follows. **a)** The $^3PF_2$ neutron pair energy gap appears, $\Delta(^3PF_2(n)) > 0$, in the region $1.3\times10^{14} < \rho$ (g/cm$^3$) $< 7.2\times10^{14}$, where $\Delta(^3PF_2(n))$ is an increasing function of the density for the region $\rho < 3.3\times10^{14}$ g/cm$^3$ and it is a rapidly decreasing function of the density for the region $\rho > 5.2\times10^{14}$ g/cm$^3$. **b)** The maximum of the $^3PF_2$ neutron pair energy gap is about 0.048 MeV at $\rho \cong 4.2$ g/cm$^3$. **c)** $\Delta(^3PF_2(n))$ is almost a constant about the maximum with error less than 3% in the rather wide range $3.3\times10^{14} < \rho$ (g/cm$^3$) $< 5.2\times10^{14}$.

The anisotropic $^3PF_2$ superfluid neutron state can exist in the rather wide density region $3.3\times10^{14} < \rho$ (g/cm$^3$) $< 5.2\times10^{14}$ when the temperature decreases down to the critical temperature, $T_\lambda$ ($^3PF_2(n)$), which corresponds to the maximum of the $^3PF_2$ neutron pair energy gap

$$T \leq T_\lambda(^3PF_2(n)) = \Delta_{\max}(^3PF_2(n))/2k \approx 2.78\times10^8 K . \qquad (1)$$

A $^3P_2$ neutron Cooper pair has a spin angular momentum with a spin quantum number, $\sigma = 1$. The $^3P_2$ neutron Cooper pair possesses a magnetic moment being twice of the abnormal magnetic moment for a neutron, $2\mu_n$ on magnitude, and its projection on an external magnetic field (z-direction) is $\mu_{\text{pair}} = -\sigma_z \times (2\mu_n)$, $\sigma_z = 1, 0, -1$.

$$\mu_n = 0.966\times10^{-23} \quad erg/gauss \qquad (2)$$

Behavior of the $^3P_2$ neutron superfluid is Similar to behavior of the liquid $^3$He in very low temperature (Leggett, 1975):

1) The projection distribution for the magnetic moment of the $^3P_2$ neutron Cooper pairs in a circumstance without external magnetic field is on equal probability, or Equal Spin Pair (ESP) phase similar to the A- phase of the liquid $^3$He in very low temperature (Leggett, 1975). The $^3P_2$ neutron superfluid is basically isotropic without significant magnetic moment in the case no external magnetic field. We name it as the A- phase of the $^3P_2$ neutron superfluid.

2) However, The projection distribution for the magnetic moment of the $^3P_2$ neutron Cooper pairs in the circumstance with external magnetic field is not on equal probability. The number of the $^3P_2$ neutron Cooper pair with paramagnetic magnetic moment is more than one with diamagnetic magnetic moment. Therefore, the $^3P_2$ neutron superfluid possesses a whole induced paramagnetic magnetic moment and its behavior is anisotropic under external magnetic field. We name it as the B- phase of the $^3P_2$ neutron superfluid similar to the B-phase of the liquid $^3$He in very low temperature (Leggett, 1975).

# III. Induced paramagnetic magnetic moment of the $^3P_2$ neutron superfluid in the B-phase

The induced paramagnetic magnetic moment of the $^3P_2$ neutron superfluid in the B-phase may be simply estimated as follows:

The system of the $^3P_2$ neutron Cooper pairs with spin quantum, s = 1, may be taken as a Bose-Einstein system. All $^3P_2$ neutron Cooper pairs may condense into same state with lowest energy (the fundamental state, E=0) in very low temperature. However, a paramagnetic magnetic moment possesses lower energy than a diamagnetic magnetic moment under the external magnetic field by stability. Therefore, the $^3P_2$ neutron Cooper pair has energy $\sigma_Z \times 2\mu_n B^{(tot)}$ ($\sigma_Z$ =1,0,-1) under the magnetic field due to its magnetic moment [$-\sigma_Z \times 2\mu_n$], where $B^{(tot)}$ is the total magnetic field including the initial background magnetic field (which originates from the collapsed core of progenitor during supernova explosion) and the induced magnetic field, $B^{(tot)} = B^{(0)} + B^{(in)}$. We denote the number of $^3P_2$ neutron Cooper pairs with spin projection $\sigma_z$ =1,0,-1 by $n_1$, $n_0$ and $n_{-1}$ respectively. Their relative ratios are

$$\frac{n_{-1}}{n_0} = e^{2\mu_n B^{(tot)}/kT}, \qquad \frac{n_1}{n_0} = e^{-2\mu_n B^{(tot)}/kT}$$

$$n_{-1} + n_0 + n_1 = n_n(^3P_2) \qquad (3)$$

The difference of the $^3P_2$ neutron Cooper pair number with paramagnetic and diamagnetic magnetic moment is

$$\Delta n_{\mp 1} = n_{-1} - n_1 \approx \frac{4}{9}\frac{\mu_n B^{(tot)}}{kT} n_n(^3P_2) \qquad (4)$$

(in a condition $\mu_n B^{(tot)} << kT$)

Neutrons combined into the $^3P_2$ Cooper pairs are just in a thin layer with thickness $\sqrt{2m_n \Delta_n(^3P_2)}$ near the surface of the Fermi sphere in the momentum sphere. The fraction of neutrons combined into the $^3P_2$ Cooper pairs is

$$q = \frac{4\pi p_F^2(n)[2m_n \Delta(^3P_2(n))]^{1/2}}{(4\pi/3)p_F^3} = 3[\frac{\Delta(^3P_2(n))}{E_F(n)}]^{1/2} \qquad (5)$$

Taken $E_F(n)$ ~ 100 MeV, $\Delta(^3P_2)$ ~ 0.05 MeV, q ~ 13%. Thus, the total number of neutrons into the $^3P_2$ Cooper pairs is

$$N_n(^3P_2 pair) \approx qN_A m(^3P_2(n))/2 \qquad (6)$$

Therefore, the total difference of the $^3P_2$ neutron Cooper pair number with paramagnetic and diamagnetic magnetic moment is

$$\Delta N_{\mp 1} = \frac{4}{9} \frac{\mu_n B^{(tot)}}{kT} N_n(^3P_2 pair) = \frac{2}{9} \frac{\mu_n B^{(tot)}}{kT} q N_A m(^3P_2(n)) \quad (7)$$

The total induced magnetic moment, of the anisotropic neutron superfluid is thus

$$\mu_{pair}^{(tot)}(^3P_2) = \Delta N_{\mp 1} \times 2\mu_n = \frac{4}{9} \frac{\mu_n B^{(tot)}}{kT} q N_A m(^3P_2(n)) \mu_n \quad (8)$$

where $m(^3P_2(n))$ is the mass of the anisotropic neutron superfluid in the neutron star, $N_A$ is the A'vogadro constant. The factor $(4/9)\mu_n B^{(tot)}/kT$ is introduced for taking the effect of thermal motion on the direction of the magnetization.

According to the formulae of the magnetic moment with the polar magnetic field $|\mu_{NS}| = B_p R_{NS}^3 / 2$ ( Shapiro and Teukolski, 1984, here $B_p$ is the polar magnetic field strength and $R_{NS}$ is the radius of the neutron star), the induced magnetic field by the total magnetic moment of the anisotropic neutron superfluid is then

$$B^{(in)} = \frac{\eta}{T_7 - \eta} B^{(0)} \quad (9)$$

$$\eta = 1.0 [\frac{m(^3P_2(n))}{0.1 m_{sun}}] R_{NS,6}^{-3} \quad (10)$$

Here $R_{NS,6}$ is the radius of the neutron star in unit of 10 km. $T_7$ is temperature in the unit of $10^7$K. The induced magnetic field of the anisotropic superfluid will increases with the interior temperature decreasing. The induced magnetic field will exceed the initial background magnetic field when the temperature decreases down to $T < 2\eta \times 10^7 K$.

For an example, for the case $\eta=3$, the induced magnetic field corresponding to the interior temperature is tabulated in the Table .

| $T_7$ | 4.0 | 3.5 | 3.3 | 3.2 | 3.1 |
|---|---|---|---|---|---|
| $B^{(in)}/B^{(0)}$ | 3.0 | 6.0 | 10.0 | 15.0 | 30.0 |

## IV. An Upper limit of the magnetic field of the neutron stars

The Formulae (4) and (7)-(9) are valid only in the condition $\mu_n B^{(tot)} \ll kT$. When the temperature decreases down to $T \sim \mu_n B^{(in)}/k$, These formulae are no longer valid. This will happen when $T_7 - \eta \approx 1.74 \times 10^{-3} \eta^2 B_{12}^{(0)}$. The corresponding formulae should be derived from eq. (3) when the temperature decreases further.

There is an upper limit of the induced magnetic field of $^3P_2$ superfluid. The upper limit will be arrived when all magnetic moments of $^3P_2$ neutron Cooper pairs are arranged with paramagnetic direction as the temperature decreases down to the absolute zero of Kelvin. The upper limit of the induced magnetic moment of $^3P_2$ superfluid is

$$\mu_{tot}^{(up)}(^3P_2) = qN_A m(^3P_2(n))\mu_n \approx 1.67 \times 10^{32} [\frac{m(^3P_2(n))}{0.1 m_{Sun}}] \quad (c.g.s.) \quad (11).$$

The upper limit of the induced magnetic field of $^3P_2$ superfluid is

$$B_{max}^{(in)} \approx 3.35 \times 10^{14} \eta \quad gauss$$

## V. Conclusion and Discussion

The conclusion of this paper is that the ultra magnetic field of the magnetars originates really by the induced magnetic field due to the induced paramagnetic moment of $^3P_2$ superfluid with significant mass more than $(0.3-0.6)m_\odot$ in the temperature lower than $3.1 \times 10^7$K.

It should be pointed out that the neutron fluid is still in the normal Fermi fluid state in the region $1.6 \times 10^{14} < \rho(\text{g/cm}^3) < 3.3 \times 10^{14}$ and $\rho > 5.2 \times 10^{14}$ g/cm$^3$ when the temperature is near the maximum transition temperature $T_\lambda(^3P_2(n))$. However, the anisotropic ($^3P_2(n)$) superfluid region will gradually enlarge when the temperature steadily decreases further, because the neutron (normal) Fermi fluid in some region will be transformed to the anisotropic ($^3P_2$) neutron superfluid when its temperature further decreases below the corresponding transition temperature, $T \leq \Delta(^3P_2(n))/2k$.. Thus both the mass of the anisotropic ($^3P_2$) neutron superfluid and the parameter, $\eta$, will increase with temperature decreasing.

For all neutron stars with a significant $^3P_2$ superfluid, their magnetic field will increase gradually as the star is cooling continually and all of them will evolution towards magnetars.

It is possible, many observed pulsars with magnetic field stronger more than $10^{12}$ gauss may be really with the strong induced magnetic field of the significant $^3P_2$ superfluid.